\documentclass[preprintnumbers,amsmath,amssymb,aps,prl,floatfix,lengthcheck,superscriptaddress]{revtex4}
\usepackage{graphicx}
\usepackage{braket}
\usepackage{upgreek}
\usepackage{color}

\newcommand{\um}{$\upmu$m}

\newcommand{\density}[2]{#1$\times$10\textsuperscript{#2}\unit{}{cm\textsuperscript{-3}}}
\newcommand{\unit}[2]{\mbox{\ensuremath{#1}\,#2}} 

\begin{document}

\title{An ionic impurity in a Bose-Einstein condensate at sub-microkelvin temperatures}

\author{K. S. Kleinbach}
\author{F. Engel}
\author{T. Dieterle}
\author{R. L\"{o}w}
\author{T. Pfau}
\author{F. Meinert}
\affiliation{5. Physikalisches Institut and Center for Integrated Quantum Science and Technology, Universit\"{a}t Stuttgart, Pfaffenwaldring 57, 70569 Stuttgart, Germany}
\date{\today}

\begin{abstract}
Rydberg atoms immersed in a Bose-Einstein condensate interact with the quantum gas via electron-atom and ion-atom interaction. To suppress the typically dominant electron-neutral interaction, Rydberg states with principal quantum number up to $n=190$ are excited from a dense and tightly trapped micron-sized condensate. This allows us to explore a regime where the Rydberg orbit exceeds the size of the atomic sample by far. In this case, a detailed lineshape analysis of the Rydberg excitation spectrum provides clear evidence for ion-atom interaction at temperatures well below a microkelvin. Our results may open up ways to enter the quantum regime of ion-atom scattering for the exploration of charged quantum impurities and associated polaron physics.
\end{abstract}

\maketitle

The level of control nowadays attained over ultracold atomic gases largely relies on the unprecedented precision with which interparticle interactions can be tuned in experiments \cite{Bloch2012,Chin2010}. While for neutral atoms the necessary regime of ultracold quantum scattering has been exploited extensively over the last decades, the situation is different for recently explored mixtures of atoms and ions due to more stringent temperature requirements \cite{Cote2016,Tomza2017}. Reaching the quantum scattering regime for these systems is expected to provide a rich experimental platform with novel phenomena and applications. Among others, those may comprise precision measurements of ion-atom collision parameters and associated molecular potentials \cite{Idziaszek2009,Schmid2017}, ultracold quantum chemistry \cite{Jachymski2013}, the study of exotic and strong-coupling impurity physics \cite{Cote2002,Schurer2017,Casteels2011}, or quantum simulations of condensed matter systems \cite{Gerritsma2012} with prospects for implementing coupling to lattice phonons \cite{Bissbort2013}.

Recent years have seen rapid progress in controlling ion-atom mixtures based on hybrid approaches by combining radio-frequency ion traps with optical traps for neutral ensembles. In such settings, cold collisions and chemical reactions have been investigated \cite{Grier2009,Ratschbacher2012,Kruekow2016,Hall2012}, including the study of single ions in Bose-Einstein condensates (BECs) \cite{Zipkes2010}. Yet, the intrinsic micromotion of the ion sets fundamental temperature limits typically in the millikelvin regime \cite{Cetina2012}, which have so far prevented reaching the elusive quantum regime. Mixtures with favorable mass ratios, however, hold promising perspectives \cite{Tomza2015,Joger2017}. Alternatively, optical trapping of ions has been recently demonstrated, but controlled mixing with ultracold atoms remains an open challenge \cite{Huber2014,Lambrecht2017}.

\begin{figure}[!ht]
\centering
	\includegraphics[width=\columnwidth]{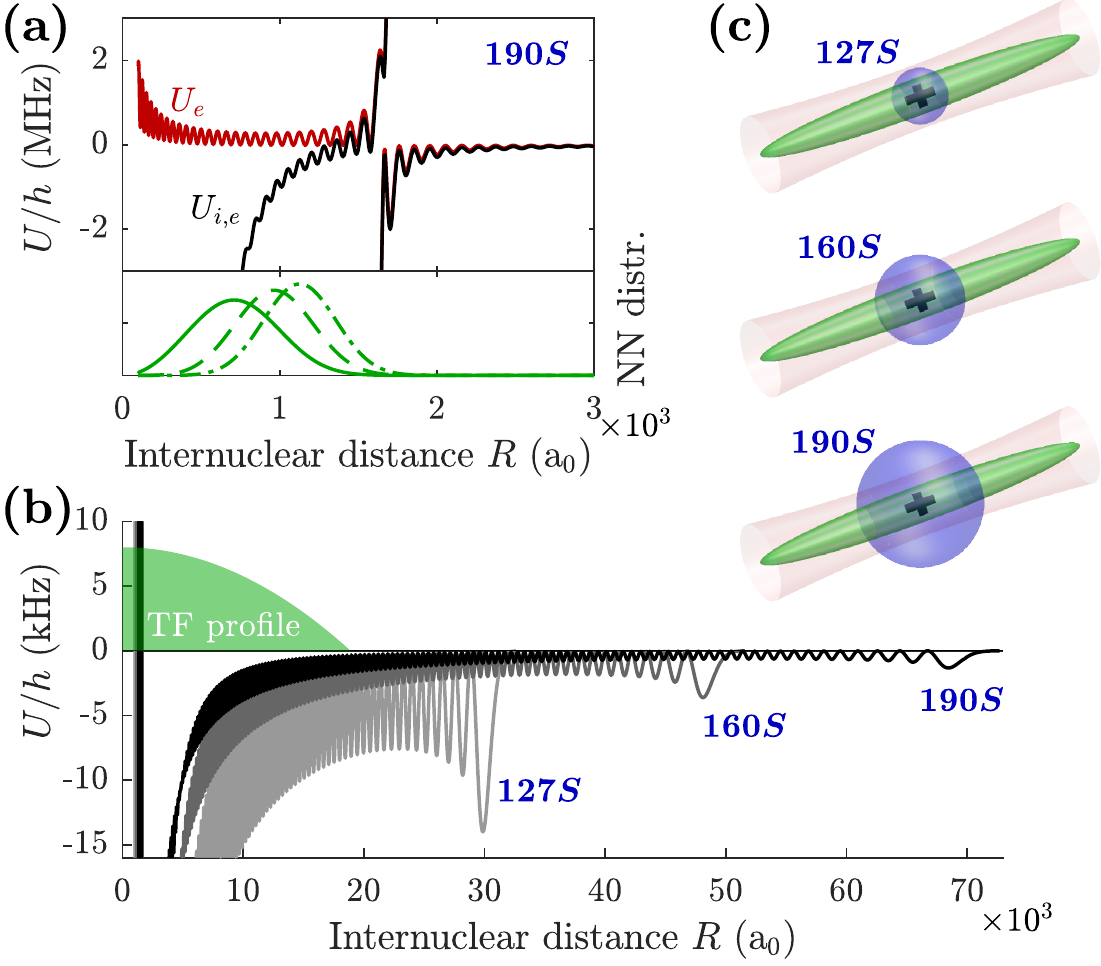}
	\caption{Concept of the experiment.
	(a) Potential energy $U_{i,e}$ (black) and $U_{e}$ (red) as a function of internuclear distance $R$ in the vicinity of the $190S_{1/2}+ 5S_{1/2}$ asymptote ($U=0$). While $U_{e}$ only accounts for the electron-neutral interaction, $U_{i,e}$ also includes the interaction with the Rydberg core ion. 
	In the lower panel, nearest (solid), next-nearest (dashed), and next-next-nearest (dash-dotted) neighbor distributions are depicted for a typical density of \density{3}{15}.
	(b) Full spatial range of $U_{i,e}$ shown for three principal quantum numbers for comparison with the spatial extent of the BEC Thomas-Fermi profile (green) along the short trap axis. 
	(c) Illustration of the BEC dimension (green), trapped in the optical tweezer (red), and the size of the $nS$ Rydberg electron orbit (blue) for the Rydberg states in (b).}
	\label{fig:Potential}
\end{figure}

In this Letter, we explore a novel approach and study the ion-atom interaction for the core of a giant Rydberg atom immersed in a BEC of $^{87}$Rb. Exciting the Rydberg state from a condensed sample conceptually maintains the ultralow temperature environment of the parent atomic ensemble, though in our experiment we are mainly limited by the imparted photon-recoil during Rydberg excitation. The interaction of the BEC with the core ion is thus probed at temperatures below a microkelvin, which is about three orders of magnitude lower compared to what has been achieved in more conventional hybrid traps \cite{Meir2016,Haerter2012}. Yet, our system temperature is still above the $s$-wave scattering limit, which for Rb is $E^{\star} \sim (2 \mu^2 C_4)^{-1}$\mbox{$= k_B \times$ \unit{79}{nK}} \cite{Grier2009,AU}. Here, $\mu$ denotes the reduced mass and $C_4 = $ \unit{318.8}{a.u.} the atomic ground-state polarizability \cite{Holmgren2010}. A striking advantage of our method is that rapid acceleration of the ionic impurity due to detrimental electric stray fields is prevented by the Rydberg electron which provides an effective shielding.

To discuss the concept of the experiment, we first consider a Rydberg atom interacting with a single neutral ground-state atom, which resides within the Rydberg electron orbital. The position $\mathbf{R}$ of the atom and $\mathbf{r}$ of the electron are measured relative to the central Rydberg ionic core. The polarizable neutral particle interacts with both the charged core ion and the electron. The first contribution is described by the classical ion-atom polarization potential
\begin{equation}
V_i= - \frac{C_4}{2 R^4} ,
\end{equation}
where $R = |\mathbf{R}|$ denotes the internuclear distance. The interaction of the low-energy Rydberg electron with the neutral atom requires a quantum mechanical formulation with Fermi's pseudopotential  \cite{Omont1977,Hamilton2002,Greene2000,Bendkowsky2009}
\begin{equation}
	V_e= 2\pi a_s(k)\ \delta^3(\mathbf{r}-\mathbf{R}) +6\pi a_p(k)\delta^3(\mathbf{r}-\mathbf{R})\overleftarrow{\nabla} \cdot \overrightarrow{\nabla} ,
\end{equation}
where the (triplet) $s$-wave, and $p$-wave scattering terms are specified by the respective energy-dependent scattering lengths $a_{s,p}(k)$. We omit the singlet scattering channels \cite{Anderson2014,Sassmannshausen2015}, which do not play a role for the electron spin configuration studied in this work. The relative strength of the ion-atom and electron-atom interaction depends on $R$, and can be quantified by solving the system Hamiltonian including both interaction terms. A full diagonalization on a truncated Hilbert space yields the two-body Born-Oppenheimer potential energy $U_{i,e}(R)$ \cite{SM,Kleinbach2017} shown in Fig.~\ref{fig:Potential} for $\ket{nS_{1/2}}$ Rydberg states.

At large internuclear distance $U_{i,e}$ is fully determined by the electron-atom interaction and inherits its characteristic shape from the Rydberg electron wavefunction as a consequence of the short-range nature of $V_{e}$ (Fig.~\ref{fig:Potential}(b)). Only when the atom is close to the Rydberg core, the ion-atom interaction starts to compete and finally dominates for sufficiently small $R$. The zoom-in depicted in Fig.~\ref{fig:Potential}(a) reveals the contribution of the ion-atom potential for \mbox{$R <$ \unit{3000}{$a_0$}} (Bohr radius) by comparing $U_{i,e}$ to $U_{e}$. For the latter, we omit $V_{i}$ in the computation of the potential energy curve. Note that the divergence at \mbox{$R$ \unit{\approx 1700}{$a_0$}} is due to a shape resonance in the electron-atom $p$-wave scattering channel \cite{Hamilton2002}.

The potential energy curve now forms the basis for the description of a single Rydberg atom interacting with a dense BEC, for which many neutral perturbers are found within the range of $U_{i,e}$. For the non-degenerate $S$-orbitals considered in this work, the interaction of each perturber contributes individually and the joint effect can be probed by analyzing interaction-induced broadenings and shifts of the Rydberg laser excitation spectrum \cite{Balewski2013,Camargo2017}. Evidently, probing the ion-atom interaction poses a two-fold challenge. First, the BEC density needs to be high enough so that sufficiently many atoms are located within the range where $V_{i}$ starts to compete with $V_{e}$. Second, the typically dominant contribution of the electron-atom interaction at large $R$ has to be decreased. The latter can be achieved by reducing the spatial overlap between the Rydberg electron wavefunction and the BEC density distribution. In our experiment, we address both aspects by confining an elongated BEC in a tightly focused optical tweezer. This provides high peak densities with a typical nearest neighbor spacing of \unit{\approx 700}{$a_0$} in the cloud center (\textit{cf.} Fig.~\ref{fig:Potential}(a)). Note that the characteristic range of $V_{i}$ for Rb is $R^{\star} = \sqrt{\mu C_4}$ \unit{\approx 5000}{$a_0$} \cite{Tomza2017}. At the same time we access a regime, where for large $n$ the Rydberg orbit reaches far beyond the radial extent of the condensate, thereby reducing the contribution of electron-atom interaction (Fig.~\ref{fig:Potential}(b) and (c)).

In a first set of experiments, we demonstrate access to the ion-atom interaction by suppressing the effect of the Rydberg electron. For this, we tune the principal quantum number up to $n=190$ (orbital radius \unit{\approx 3.7}{\um}). Our measurements start from a BEC of typically $6 \times 10^5$ $^{87}$Rb atoms in the $\ket{5S_{1/2},F=2,m_F=2}$ hyperfine state, prepared in a magnetic Quadrupole-Ioffe-Configuration (QUIC) trap at a temperature below \unit{250}{nK}. A small volume optical tweezer, focused by a high-NA aspheric lens (Gaussian waist \unit{\approx 1.8}{\um}, wavelength \unit{855}{nm}) and overlapped with the condensate, is loaded within \unit{10}{ms}. Subsequently, a small change of the magnetic trapping fields allows us to shift the parent BEC aside in order to isolate the optically trapped sample. At the end of the preparation procedure, we achieve micron-sized elongated BECs of typically $6.5 \times 10^4$ atoms and peak densities $\approx$\density{3}{15}. Radial and longitudinal trap frequencies for the confined sample are measured to \mbox{$\omega_\perp = 2 \pi \times$ \unit{2180(60)}{Hz}} and \mbox{$\omega_{||}=2 \pi \times \unit{215(30)}{Hz}$}, respectively.

\begin{figure}[!th]
\centering
	\includegraphics[width=\columnwidth]{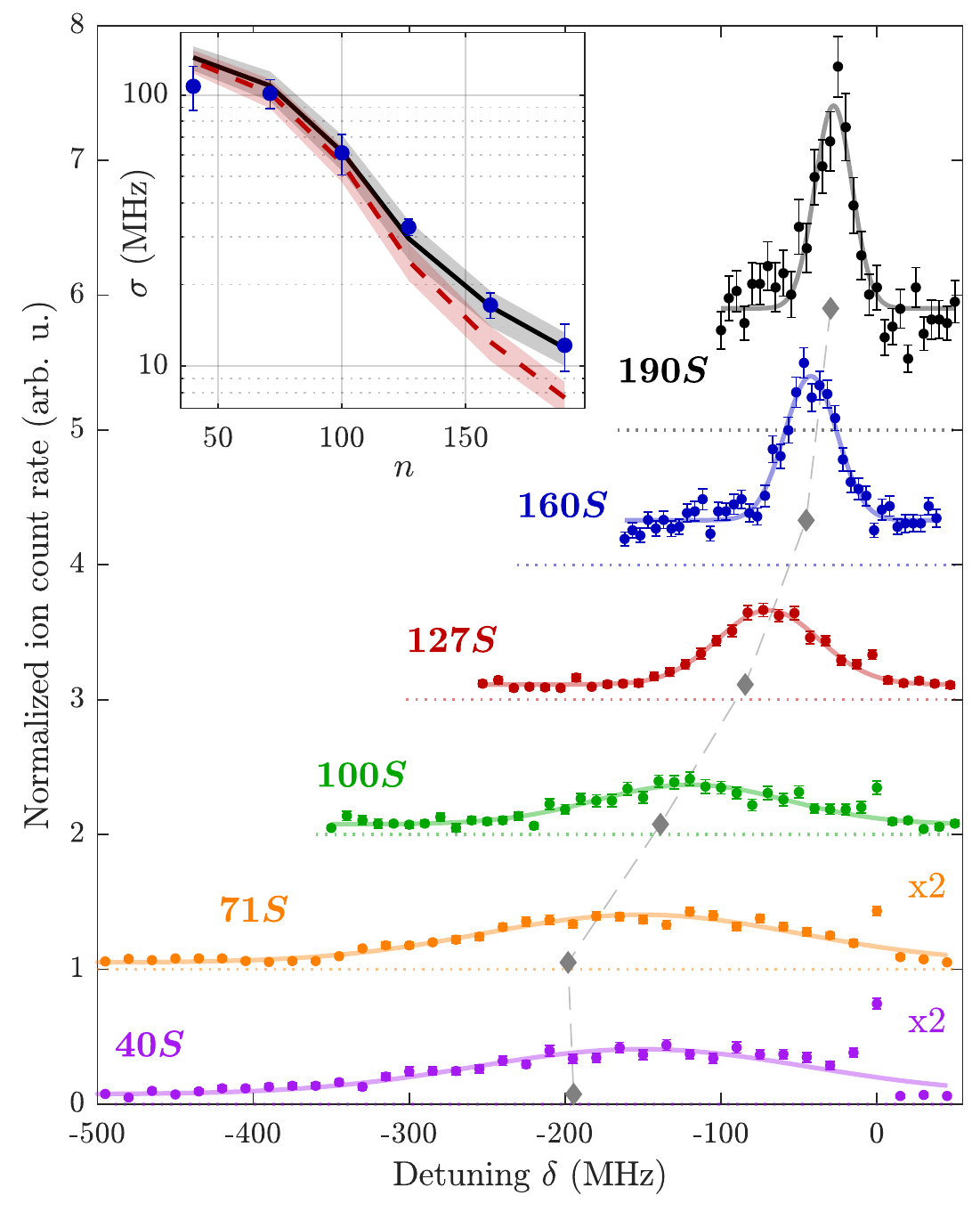}
	\caption{
	Rydberg spectroscopy in the BEC. The normalized ion count rate is shown as a function of laser detuning $\delta$ with respect to the bare $\ket{nS_{1/2}}$ Rydberg resonance ($\delta=0$) for a set of principal quantum numbers $n$ as indicated. Solid lines are Gaussian fits to the data to extract the spectral width $\sigma$. The datasets are offset for better readability and zero count rate is denoted by the dotted lines. The data for $n=40,71$ is scaled by a factor of two. Error bars show $1\sigma$ statistical uncertainty. The filled diamonds indicate the center of the excitation spectra predicted from a full numerical simulation (see text). Inset: Spectral width $\sigma$ as a function of principal quantum number. Error bars indicate the confidence interval from the fitting procedure to extract $\sigma$. The solid black (dashed red) line shows the prediction from our numerical simulation with (without) taking the ion-atom interaction into account. Shaded regions indicate the experimental uncertainty in atom number ($\pm10\%$) and trapping parameters.
	}
	\label{fig:SpectraWaterfall}
\end{figure}

In the dense micro-BEC, we now excite a single atom to the $\ket{nS_{1/2},m_J=+1/2}$ Rydberg state via two-photon excitation incorporating the intermediate $6P_{3/2}$ level at a detuning of \unit{+80}{MHz}. For this, the sample is illuminated simultaneously with two frequency-tuneable laser beams at wavelengths \unit{420}{nm} and \unit{1020}{nm}. The latter is focused through the same lens that is used to generate the tweezer trap to ensure local Rydberg excitation only in the micro-BEC \cite{Schlagmueller2016b}. For the bare Rydberg atom, the excitation scheme transfers a photon-recoil energy of \mbox{$k_B \times $\unit{730}{nK}}. Each excitation pulse of \unit{500}{ns} (\unit{200}{ns} for $n=40$ and $71$) is followed by field-ionization and detection (efficiency $>40\%$) of the produced ion on a microchannel plate detector. This procedure is repeated five times with a repetition rate of \unit{20}{kHz} in the same atomic sample. To avoid any Rydberg-Rydberg interaction, the ion count rate is kept well below one ($<0.3$ ions/pulse). Rydberg spectra are obtained by variation of the two-photon detuning $\delta$ and averaging over at least 75 realizations. For high-$n$ Rydberg states ($n\geq127$) we account for diamagnetic line shifts that arise due to the magnetic field ramps to a final value of \unit{7.73}{G} during sample preparation \cite{SM}. 

Results of such measurements for increasing values of $n$ are shown in Fig.~\ref{fig:SpectraWaterfall}. For all datasets, we observe large spectral redshifts and strong line broadening \cite{SM}, which are attributed to the interaction of the Rydberg atom with the BEC. While shift and broadening is comparable for $40S_{1/2}$ and $71S_{1/2}$, the spectrum narrows and shifts towards the bare Rydberg transition at $\delta=0$ when $n$ is further increased.  The rather independent spectral shape when changing $n$ from $40$ to $71$ is indeed expected as long as the electron orbit is considerably smaller than the condensate, which has been experimentally confirmed in earlier studies \cite{Balewski2013,Camargo2017,Schlagmueller2016b}. In this regime, the spectrum is dominated by the electron-neutral interaction. For larger $n$, the decreasing shift and broadening is due to the reduced overlap of the electron with the BEC, and consequently the suppressed electron-atom interaction.

To quantify the effect of the ion-atom interaction, we compare our measurements to a full numerical simulation of the measured spectral lineshape. Our theoretical analysis starts from the two-body potential energy curves $U_{i,e}$ and $U_{e}$ as introduced above. We recall that $U_{i,e}$ takes into account both, the ion-atom as well as the electron-neutral interaction, whereas the former is omitted in $U_{e}$. For modelling the presence of many ground-state perturbers, we apply a Monte-Carlo sampling approach which treats the atoms from the BEC as point-like particles that are randomly distributed within the range of $U$ according to the condensate density distribution in the trap. The model also takes into account the random position of the ionic core in the condensate weighted by the Rydberg excitation laser profile. From each realization a density-induced energy shift is calculated. Averaging over typically $5 \times 10^5$ realizations and taking into account the finite excitation linewidth of the unperturbed Rydberg atom delivers a precise lineshape for the excitation spectrum. Note that this model has been previously applied to analyze the electron-neutral scattering in regimes, where the ion-atom interaction is negligible \cite{Schlagmueller2016b,Camargo2017}. Moreover, our semi-classical sampling method has recently been shown to reproduce a full quantum mechanical treatment based on a functional determinant approach at sufficiently large densities \cite{Schmidt2016,Schmidt2017}. We have verified that this holds for our system parameters and also when including the ion-atom interaction \cite{SchmidtPC}.

The ion-atom interaction is most clearly identified when analyzing the width of the excitation spectrum. The measured width $\sigma$, extracted from Gaussian fits to the data, is shown as a function of $n$ in the inset to Fig.~\ref{fig:SpectraWaterfall}. In our fitting procedure, we account for an overall offset in the data, which is more prominent for higher $n$. Partly, this offset can be attributed to direct photoionization of atoms by the tweezer light in combination with the \unit{420}{nm} Rydberg excitation laser. The experimental results are compared to the width extracted from our numerical simulation based on $U_{i,e}$ (black line) and $U_{e}$ (red line). The interaction of the ion with the BEC is evident for large $n$ and causes an increased width of the excitation spectrum in accordance with our measurements. Additionally, we compare the center position of the simulated spectra using $U_{i,e}$ (diamonds in Fig.~\ref{fig:SpectraWaterfall}) with the data  and find good agreement over the entire range of investigated principal quantum numbers. Note that the ion-atom interaction also causes a redshift of the center position (\unit{\approx 5}{MHz}) for large $n$, which is yet less prominent than the effect on the spectral width.

\begin{figure}[!t]
\centering
	\includegraphics[width=\columnwidth]{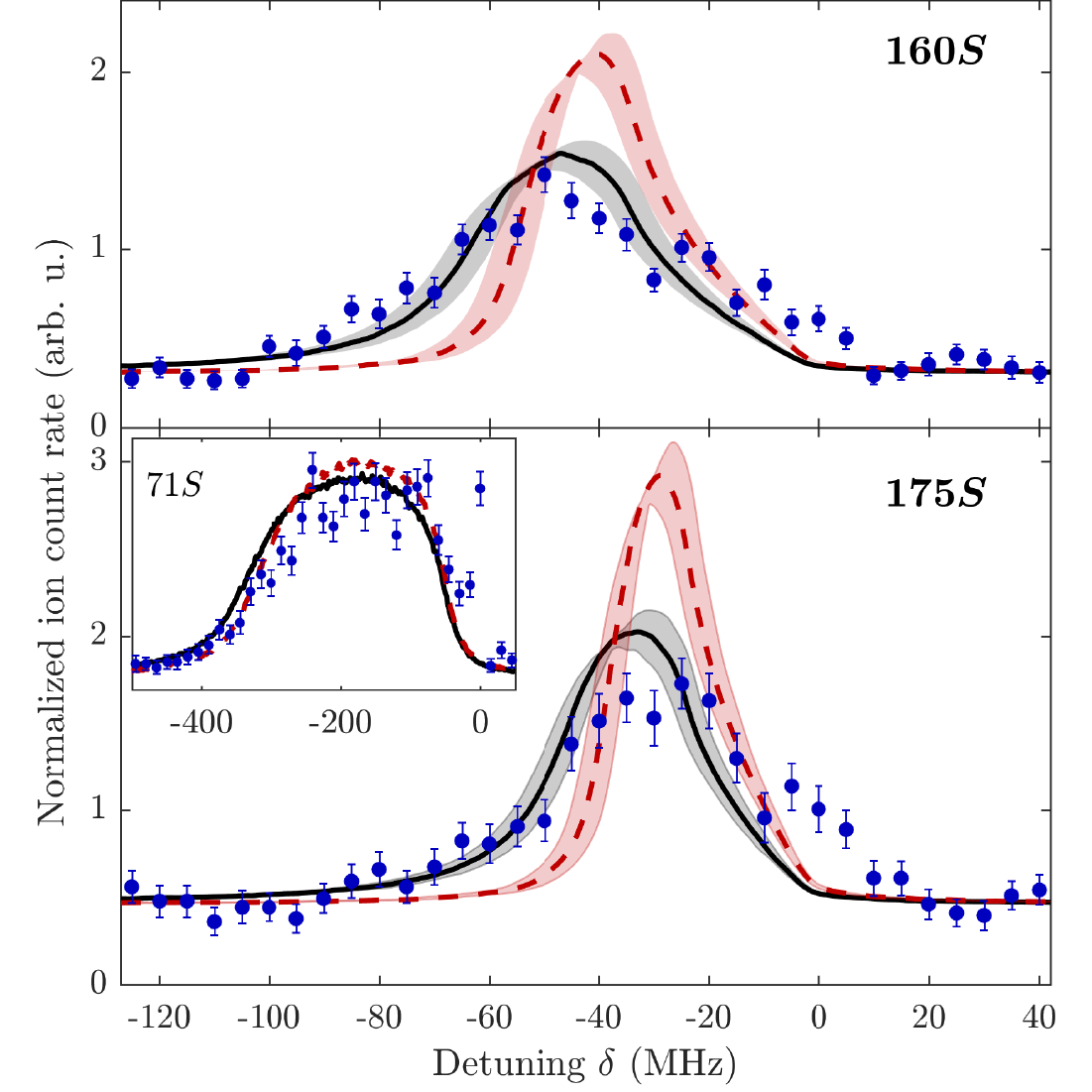}
	\caption{
	Contribution of ion-atom interaction to Rydberg spectra for high-$n$ Rydberg states ($n=160,\ 175$; inset $n=71$). The normalized ion count rate is shown as a function of laser detuning $\delta$ in the vicinity of the bare $\ket{nS_{1/2}}$ Rydberg resonance ($\delta=0$). The solid black (dashed red) line shows the result of our full numerical lineshape simulation with (without) taking the ion-atom interaction into account. The shaded areas indicate experimental uncertainty dominated by a $\pm10\%$ error on the atom number ($N=4.8 \times 10^4$ ($71S$); $N=5.1 \times 10^4$ ($160S$); $N=4.6 \times 10^4$ ($175S$)). Error bars show $1\sigma$ statistical uncertainty.
	}
	\label{fig:CompareC4}
\end{figure}

Next, we elaborate on the full lineshape of the excitation spectra for large $n$ in order to provide a more detailed analysis of the effect of the ion-atom interaction. For this, we focus on measurements at $n=160$ and $175$ in a second set of experiments, and take explicit care to ensure an improved accuracy in determining all relevant system parameters. More specifically, the loading sequence of the tweezer trap is slightly altered in order to ensure adiabaticity when separating the micro-BEC from the parent condensate. In combination with an improved trap alignment this provides a more precise determination of the trap frequencies measured to \mbox{$\omega_{\perp} = 2 \pi \times $\unit{2442(20)}{Hz}} and \mbox{$\omega_{||} = 2 \pi \times $\unit{268(4)}{Hz}}, respectively. At the end of the sample preparation, the residual magnetic field is \unit{1.74}{G}. Additionally, the experimental sequence is repeated at least 100 times for each datapoint to reduce statistical uncertainty.

The results of these measurements are shown in Fig.~\ref{fig:CompareC4} and compared to the numerically simulated spectral profiles within our theoretical analysis outlined above. We stress that the BEC parameters entering the computed spectra are determined via independent measurements. Consequently, there are no free parameters except for the area under the curves, which is normalized for comparison to the experiment. Evidently, when we omit the ion-atom interaction and only include $U_{e}$ in the model (red dashed line), the simulation results clearly fail to capture the experimental data. Specifically, the measured spectra extend significantly more towards larger red detuning as a consequence of the presence of the ion. Indeed, taking into account the ion-atom interaction, i.e. using $U_{i,e}$ in the numerical simulation (black solid line), we find good agreement between experiment and theory. Deviations at very small detunings are possibly due to a residual thermal fraction with reduced density. Our experimental results thus provide evidence for the role of ion-atom interaction for a BEC within a giant Rydberg orbit. We stress again the importance to control Rydberg orbitals that reach far beyond the condensate dimension in order to suppress the contribution of electron-neutral scattering. This is exemplified in the inset to Fig.~\ref{fig:CompareC4} for data taken at $n=71$, where the simulations clearly indicate the dominant role of the Rydberg electron. 

\begin{figure}[!b]
\centering
	\includegraphics[width=\columnwidth]{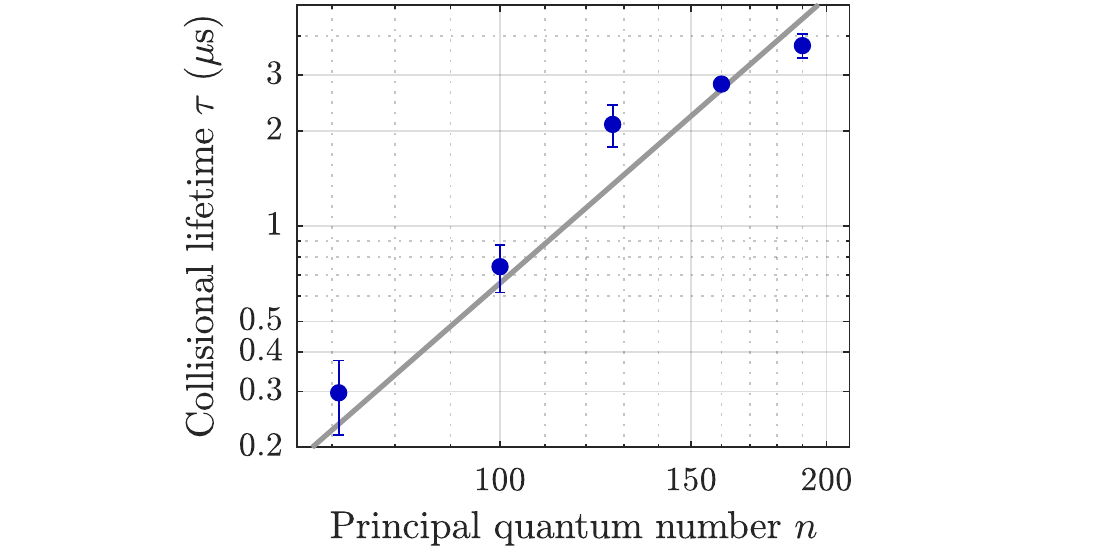}
	\caption{
	Collisional lifetime $\tau$ of the Rydberg excitation in the BEC as a function of principal quantum number $n$. The solid line is a fit to the data based on a $\sim n^3$ scaling.
	}
	\label{fig:Lifetime}
\end{figure}

Finally, we report measurements of the Rydberg atom lifetime in the dense media. For this, $\delta$ is fixed for each $n$ to the respective center of the spectrum extracted from the Gaussian fits in Fig.~\ref{fig:SpectraWaterfall}. The lifetime is then measured using state-selective field-ionization \cite{SM,Schlagmueller2016}. We idenfity state changing collisions from the initial $S$-state into high-$L$ Rydberg states involving a ground-state atom as the dominant decay channel \cite{Schlagmueller2016}. Consequently, this process causes loss of our ionic impurity due to kinetic energy release. The collisional lifetime $\tau$ as a function of $n$ is shown in Fig.~\ref{fig:Lifetime} and reveals a strong dependence on the principal quantum number. A similar trend has been also observed in a BEC which was larger than the Rydberg orbit \cite{Schlagmueller2016}. This suggests that the lifetime-limiting collisions are triggered by the simultaneous presence of the Rydberg electron and a ground-state atom in the vicinity of the core ion. In this view, an empirical $\sim n^{3}$ scaling is fit to the data (gray line), reminiscent of the variation of the Rydberg $nS$ electron density close to the ion.

In conclusion, we have explored a novel method to probe charge-neutral interaction for a single ionic impurity immersed in a condensed Bose gas at temperatures below a microkelvin. For this, we exploit the core of a Rydberg atom whose electron is located predominantly outside the condensate and protects the ion impurity from electric stray fields. The ion-atom interaction is accessed via high-resolution Rydberg spectroscopy. Indeed, our method shows conceptual similarities with ZEKE spectroscopy of molecular ions within high-$n$ Rydberg orbitals \cite{Mueller1998,Beyer2016}, but is here applied in the context of an ultracold many-body system. We believe that our work opens new possibilities to enter the quantum scattering regime for ultracold ion-atom systems. Specifically, the exploration of long-lived circular Rydberg states holds promising perspectives \cite{Anderson2013,Dunning2010,Signoles2017}. Those may not only further reduce the electron-neutral interaction but should also boost the system lifetime by orders of magnitude. This could allow for reaching motional timescales of the ion, possibly extended by optical trapping techniques \cite{Anderson2011}, to open new routes to explore many-body polaron physics.

We thank R. Schmidt and T. Schmid for critical reading of the manuscript, K. Jachymski, C. Fey, P. Schmelcher, and F. Merkt for fruitful discussions, and I. I. Fabrikant for providing scattering phase shifts.
We acknowledge support from Deutsche Forschungsgemeinschaft [Projects No. PF 381/13-1 and No. PF 381/17-1, the latter being part of the SPP 1929 (GiRyd)].
F. M. acknowledges support from the Carl-Zeiss foundation and is indebted to the Baden-W\"urttemberg-Stiftung for the finanical support by the Eliteprogramm for Postdocs.

\clearpage

\section{Supplementary Material: An ionic impurity in a Bose-Einstein condensate at sub-microkelvin temperatures}

\subsection{Diamagnetic shift for high $n$ Rydberg states \\and electric field compensation}

For the data shown in Fig.~2 of the main article, the position and width of the bare Rydberg transition ($\delta=0$) is calibrated in a dilute thermal sample. For sufficiently long excitation pulses, the linewidth is limited by electric stray fields. Careful electric field compensation allows us to achieve Gaussian widths of the bare Rydberg transition which are $<$\unit{1}{MHz} for $n$ up to $175$ and $<$\unit{3}{MHz} for $n=190$. We have verified, that this broadening does not significantly affect our calculated spectra.

Due to the magnetic field ramp applied for isolating the micro-BEC from the parent condensate, the magnetic offset field present during calibration differs from the field at which the spectra in Fig.~2 have been taken. In the presented spectra, the resulting diamagnetic lineshifts are calibrated and corrected for.

\begin{figure}[!b]
\centering
	\includegraphics[width=\columnwidth]{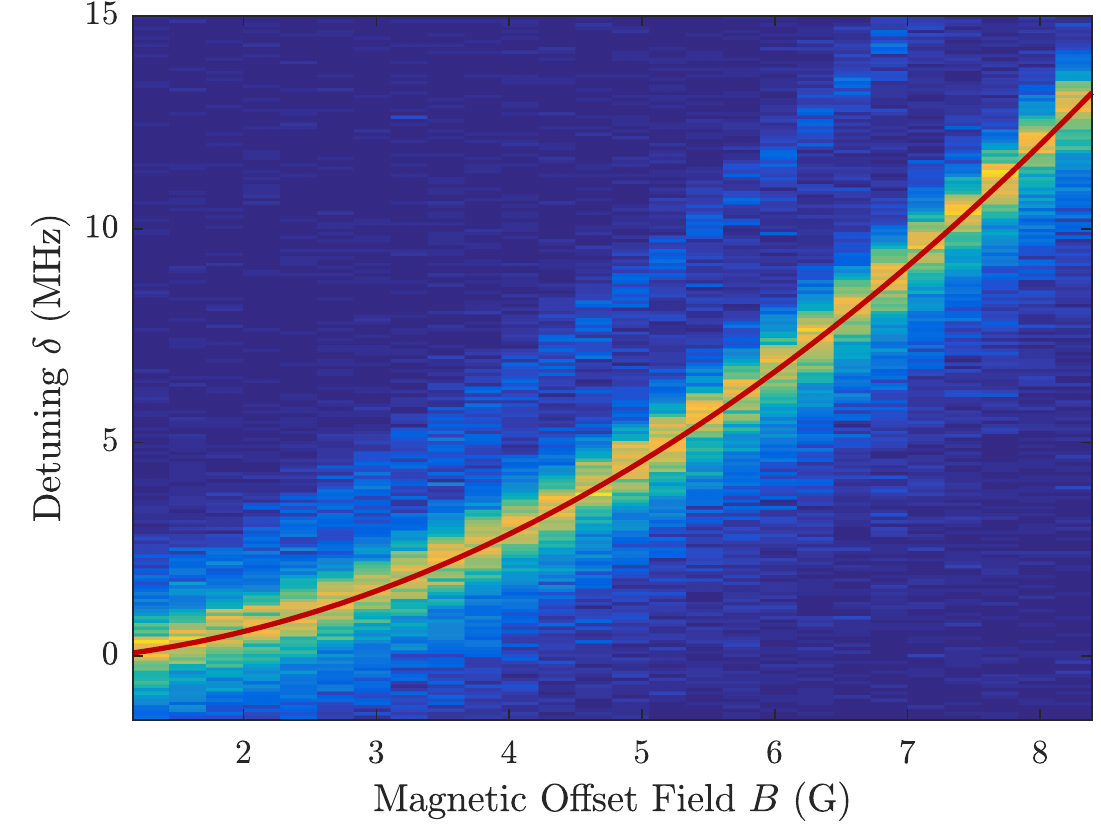}
	\caption{
Rydberg spectra for the $160S_{1/2}$ Rydberg state at varying magnetic offset fields $B$. Zero detuning is referenced to the atomic Rydberg resonance at zero magnetic field. The red line shows a fit to the data based on Eq.~\ref{Eq:Diamagnetism} to the extracted line centers.
	}
	\label{fig:Diamag}
\end{figure}
Specifically, for a Rydberg atom in a magnetic field of strength $n^4B>1$ the diamagnetic interaction gets stronger than the linear Zeeman term. As we investigate Rydberg atoms with $n>150$, this regime is entered already at a relatively small magnetic field $B<$ \unit{10}{G}. For $nS$-states in a sufficiently low magnetic field $B$ oriented along the $z$-axis the diamagnetic energy shift of the Rydberg level is given by \cite{Gallagher1994MAT}
\begin{equation}
	\Delta E_{dia} = \frac{B^2}{8} \bra{nL_1m_{L_1}}r^2sin(\theta)\ket{nL_1m_{L_1}},
\label{Eq:Diamagnetism}
\end{equation}
which scales as $n^4$. Here, $r$ is the radial position of the Ryd\-berg electron relative to the ionic core and $\theta$ its angle with respect to the $z$-axis. The Rydberg state is described by the principal quantum number $n$, the orbital quantum number $L_1$ and its projection onto the $z$-axis $m_{L_1}$. 
 
Our atomic sample is prepared in the $\ket{5S_{1/2},F=2,m_F=2}$ state (Land\'{e} factor $g_F=1/2$), which features the same linear Zeeman shift as the $\ket{nS_{1/2},m_J=1/2}$ Rydberg states (Land\'{e} factor $g_J=2$). Hence, for the optical transition, the linear Zeeman effect cancels and only the diamagnetic term remains. To quantify the diamagnetic shift, Rydberg spectra with $n\geq127$ have been measured for varying magnetic offset fields of our QUIC trap in a dilute sample. An exemplary dataset for the $160S_{1/2}$ state is shown in Fig.~\ref{fig:Diamag}. From a fit to the data based on Eq.~\ref{Eq:Diamagnetism}, we obtain a correction of \unit{11.8}{MHz} for the magnetic field ($B=$ \unit{7.73}{G}) present during Rydberg spectroscopy in the micro-BEC. The same procedure has been applied to the data for $n=127$ and $n=190$ in Fig.~2, resulting in shifts of \unit{2.9}{MHz} and \unit{22.1}{MHz}, respectively.

Note that for the data shown in Fig.~4 of the main article, the altered loading procedure of the micro-BEC results in a negligible change of the magnetic field strength, and consequently no significant diamagnetic shift.

\subsection{Measurement of the collisional lifetimes $\tau$}

For the measurement of the Rydberg atom's collisional lifetime, we apply state-selective field ionization, mainly following the procedure described in Ref.~\cite{Schlagmueller2016MAT}. Therein, it was observed, that the Rydberg lifetime in a high density and ultracold environment is limited by two processes: $L$-changing collisions of the Rydberg electron and associative ionization forming Rb$_2^+$ molecular ions. For high principal quantum numbers ($n \gtrsim 90$) the first process dominates.

In order to discriminate the initial $S$-state from the high-$L$ product state, we exploit their different ionization thresholds. Specifically, $S$-states tend to ionize adiabatically with a threshold close to the classical limit of $1/(16n^4)$. In contrast, the high-$L$ states ionize diabatically at higher field strength $\sim 1/(9n^4)$ \cite{Gallagher1994MAT}. In the experiment, we wait for a variable delay time $t$ after Rydberg excitation, and subsequently apply a two-step ionization sequence. The first part ionizes predominantly $S$-states, while the second part ionizes all remaining Rydberg atoms, including high-$L$ states. The two pulses are sufficiently delayed in time, so that the ionization products can be distinguished by their arrival time on the detector. Additionally, associated Rb$_2^+$ ions are also distinguished via time-of-flight due to their higher mass. We determine the fraction of detected $S$-states $p_{S}$ and the sum of measured high-$L$ and Rb$_2^+$ signal ($1-p_{S}$). This routine is repeated for increasing ionization delay times $t$. 

An exemplary dataset for $n=160$ is shown in Fig.~\ref{fig:Lifetime}(a). We extract the Rydberg lifetime by fitting an exponential decay according to
\begin{equation}
	p_{S}(t) = (1-c)+c\cdot \exp\left(-\frac{t-t_0}{\tau}\right).
	\label{eq:Lifetimes}
\end{equation}
Here, $c$ and $t_0$ are constants accounting for finite discriminability and pulse lengths.

\begin{figure}[!t]
\centering
	\includegraphics[width=\columnwidth]{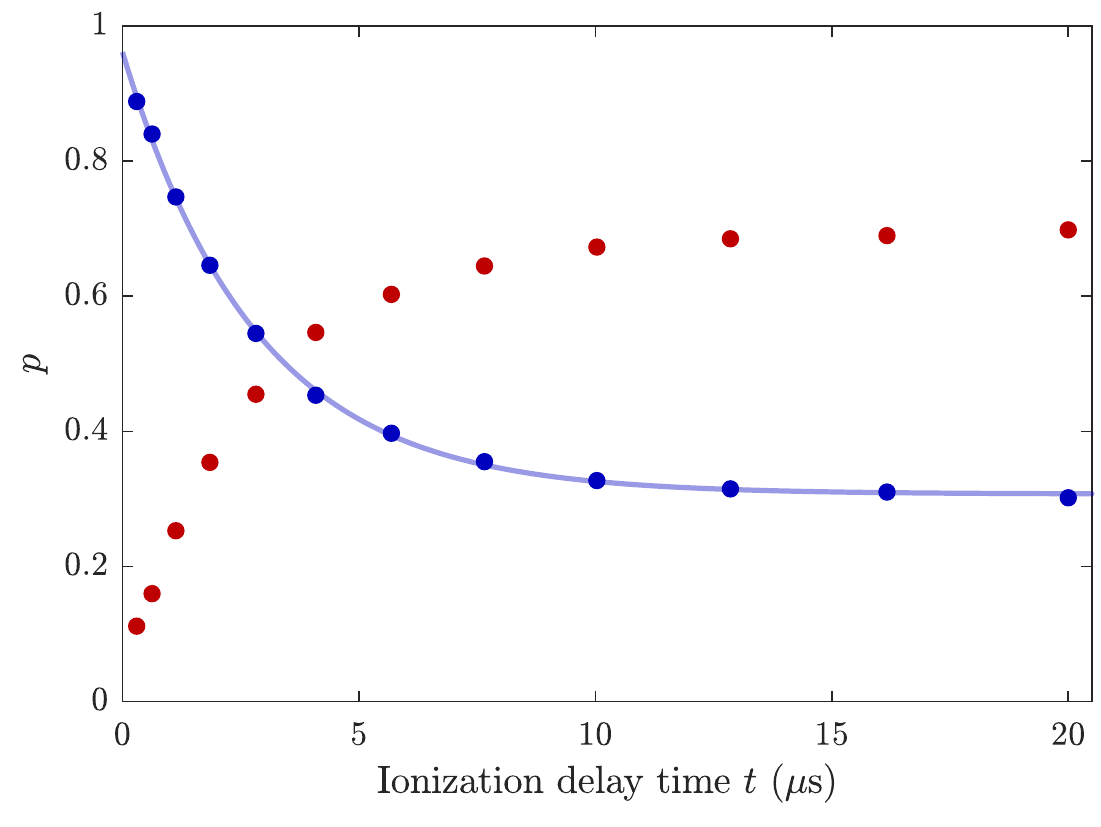}
	\caption{
	Fraction of $S$-states $p_{S}$ (blue symbols) ionized by the first field pulse (\unit{0.85}{V/cm}) and fraction of high-$L$ states (red symbols) ionized by the second field pulse (\unit{3.6}{V/cm}) of the state-selective ionization sequence as a function of delay time $t$. The data is taken for the $160S$ state at a detuning $\delta=$ \unit{42.2}{MHz}, corresponding to the line center of the spectrum in Fig.~2. Statistical error bars are smaller than the symbol size. The blue solid line is a fit to the data based on Eq.~\ref{eq:Lifetimes}.
	}
	\label{fig:Lifetime}
\end{figure} 

Note that for the Rydberg spectra, presented in Fig.~2 and 4 of the main article, the ionization fields are chosen large enough to ionize both, $S$-states and high-$L$ states, and are switched on after a fixed delay time of \unit{200}{ns}.
\subsection{Numerical calculation of the potential energy curves $U$}

For the calculation of the Born-Oppenheimer potential energy curves $U_{i,e}$, the matrix elements of the Hamiltonian
\begin{align}
	\hat{H} =  -\frac{C_4}{2\mathbf{R}^4} + \hat{H}_0 & + 2\pi a_s(k)\ \delta^3(\mathbf{r}-\mathbf{R})  \nonumber  \\ 
                &+6\pi a_p(k)\delta^3(\mathbf{r}-\mathbf{R})\overleftarrow{\nabla} \cdot \overrightarrow{\nabla}
	\label{eq:Hamiltonian}
\end{align}
are evaluated for fixed values of $R$ in a basis set $\ket{n,L_1,J_1,m_{J_1};m_{S_2}}$ \cite{Greene2000MAT,Fabrikant2002MAT,Hamilton2002MAT}. Here, $\hat{H}_0$ denotes the Hamiltonian of the unperturbed Rydberg electron including fine structure coupling between the orbital angular momentum $\hat{\mathbf{L}}_1$ and the spin angular momentum $\hat{\mathbf{S}}_1$ using published quantum defects \cite{Mack2011MAT,Li2003MAT,Han2006MAT}. The Rydberg electron states are specified by their principal quantum number $n$, their orbital angular momentum $L_1$, and their total spin-orbit coupled angular momentum  $J_1$ with its projection $m_{J_1}$ along $\mathbf{z}$. The projection of the electronic spin of the ground-state perturber is denoted with $m_{S_2}$. Note that for the spin configuration studied in this work, there is no coupling to the singlet electron-neutral scattering channel and consequently, the nuclear spin of the ground-state perturber does not play a role. The energy-dependent scattering lengths relate to the corresponding phase shifts $\delta_{s,p}(k)$ via $a_s(k)  = -\tan(\delta_s(k))/k$ and $a_p(k)  = -\tan(\delta_p(k))/k^3$ and are taken from Ref.~\cite{Fabrikant1986MAT}. For the $R$-dependent electron momentum $k$ in the scattering process, we use the semi-classical expression for the kinetic energy of the Rydberg electron $k(R)^2/2 = -1/(2 n^{\star 2}) + 1/R$, where $ n^\star$ is the effective principal quantum number of the Rydberg level of interest \cite{Greene2000MAT}. In the first term of Eq.~\ref{eq:Hamiltonian}, which accounts for the ion-atom interaction, we use the measured atomic polarizability $\alpha=C_4=$\unit{318.8}{a.u.} reported in Ref.~\cite{Holmgren2010MAT}.

The potential energy curves $U_{i,e}$ are finally obtained by full diagonalization of $\hat{H}$ for each value of $R$ on a finite basis set which spans two hydrogenic manifolds. The values of $R$ are spaced quadratically. $U_{i,e}$ is computed for about $n \times 18$ values of $R$ to provide an adequate resolution for the potential wells. For the calculation of $U_{e}$ we omit the first term in Eq.~\ref{eq:Hamiltonian}.

Note that spin-orbit interaction in the electron-atom $p$-wave scattering channel is not included in the Hamiltonian Eq.~\ref{eq:Hamiltonian}. For the spin configuration studied in this work, the main effect of spin-orbit coupling are slight shifts of the divergence from the $p$-wave shape resonance and small modifications of the Born-Oppenheimer potential close by. We have checked at the example of $n=160$ that including spin-orbit coupling for $p$-wave scattering does not significantly affect the simulated excitation spectrum.

\subsection{Monte-Carlo modeling of the Rydberg spectra}

For modeling the lineshape of the Rydberg spectra, we use a Monte-Carlo sampling method, which incorporates the density distribution of the condensate, the intensity profile of the excitation laser beams, and the Fourier-limited Rydberg excitation bandwidth. We start from a random configuration of atoms, reflecting the Thomas-Fermi density distribution of our micro-BEC using measured atom numbers and trap frequencies. Typical Thomas-Fermi radii are about \unit{1.0}{\um} and \unit{9.2}{\um} in the radial and longitudinal direction, respectively. The atoms are treated as point-like particles with infinite mass, neglecting any excitation dynamics. Furthermore, we assume uncorrelated atoms, as expected for a weakly interacting BEC. One of the atoms is designated to carry the single Rydberg excitation. For each of the remaining atoms, the potential energy $u_i$ is extracted from the interaction potential $U$ according to its distance $R$ to the Rydberg ionic core. The sum over the $u_i$ delivers the energy shift $U_n$ for a single Monte-Carlo configuration. Finally, the spectrum is obtained from the contribution of all $U_n$, weighted by the local excitation probability of the corresponding Rydberg atom and convoluted with a Lorentzian profile reflecting the Rydberg excitation bandwidth. Note that the beam profile of the excitation laser has minor influence on the spectral shape due to the small sample size. The area below the spectrum is finally normalized for comparison to the experimental data. 

Note that for the high-$n$ Rydberg states studied here, the averaged modification of the Rydberg $S$-orbit by one perturber is very small. For example, at $n=160$ it amounts to $\int (1-p_s)\rho\ d^3r\approx 2\times 10^{-5}$, with $p_s$ being the $S$-character and $\rho$ the normalized BEC density distribution. The $S$-character is obtained from the diagonalization procedure that yields the Born-Oppenheimer potential energy curve. The fact that this modification is small allows us to employ the pairwise interaction potential for our numerical simulations.


\begin{references}
\bibitem{AU} We use atomic units throughout this work.
\end{references}
\end{document}